\documentstyle[psfig]{camera}

\begin{document}

   
\title{ATOMIC ANTENNA MECHANISM IN HHG AND ATI}


\author{M. Yu. Kuchiev$^{(a)}$, V. N. Ostrovsky$^{(b)}$}
%
\organization
{$^{(a)}$ The University of New South Wales, Sydney, Australia\\
 $^{(b)}$ Institute of Physics, The University of St Petersburg, Russia}

\maketitle


\section{Introduction}
\label {intro}
This paper reviews recent development of the {\it atomic antenna}, 
a theoretical framework which describes a number  
of laser-induced multiphoton phenomena
in atoms.
The localization of atomic electrons inside an atom
drastically suppresses their interaction 
with a laser field.
For many processes this circumstance favors multistep
mechanisms when at first one of atomic electrons is
released from an atom by the field.
After that an interaction of the ejected electron with
the laser field results in absorption of energy 
from the field and its accumulation in the form of the electron 
wiggling energy and ATI energy.
It is very essential that the energy absorbed
by the electron can be 
transferred to the parent atomic core via an inelastic collision 
of the primarily ejected electron with the atom. 
The collision may trigger a number of phenomena including 
high harmonic generation  
(HHG),
enhancement of 
above-threshold ionization (ATI), production of multiply charged ions.
In this physical picture the absorption of energy from
the field takes place in the region of large separations
from an atom, where the electron-laser interaction
dominates over the electron-core potential.
This circumstance
results in dramatic enhancement of
the probability of multiphoton processes.
Such a scenario of photoabsorption 
was suggested, apparently for the first time, 
in Ref.~\cite{Ku87}. Later the idea was rediscovered
by several authors in different contexts 
\cite{C93,KulanderA,KulanderB}.
In current literature the above described
sequence of events
is often referred to as the rescattering, the three-step mechanism,
or even the simpleman model.
The term {\it atomic antenna}, suggested in Ref.~\cite{Ku87},
refers to the fact that the firstly emitted electron plays a role 
similar to an aerial in conventional radio devices, 
enhancing the absorption.

It is very important that the physical picture
drawn above
can be implemented not only as a model, but also as
a clear and rigorous quantum formalism.
In this paper we outline two convenient ways to
implement
the atomic antenna idea.
The one originating from Ref.~\cite{Ku87}
is called the factorization technique.
It was formulated
in detail in Ref.~\cite{K95} and
recently applied to HG in Refs.~\cite{KO98let,KO99pap}.
We discuss also another, complimentary
technique, which
is referred to below as
the method of effective ATI channels.
It is close in spirit to the approach developed previously
by Lewenstein {\it et al}\/ \cite{Lew,Lewphase} and can in 
turn be linked to the Corcum model \cite{C93}.
From the first glance, these two schemes differ very significantly. 
However, we prove their identity by demonstrating that they
describe the same physical idea from different points of view.

The paper is organized as follows.
Section \ref{fact} describes the 
factorization technique which  allows one to present
the amplitude of the "complicated"
multiphoton process as a product of the amplitudes
of much more simple, "elementary" processes.
Section \ref{eff} is devoted to
the concept of effective ATI channels, which 
provides important insights into the physical nature
of complicated multiphoton processes.
We derive the effective ATI channels from the factorization
technique revealing close links between the two approaches.
Section \ref{ele} describes two important examples of
"elementary" processes: one of them is the photoionization,
another one is the electron-atom collision in a laser
field which results in the generation of the high-energy quanta.
These two "elementary" processes are vital for description of 
HG in the framework of the factorization technique.
Sections \ref{hhg}, \ref{ati} are devoted to several
examples illustrating numerically an accuracy of
methods developed for HHG and ATI.
A number of results reported in this paper are derived
neglecting the Coulomb field of the residual atomic particle
which influence the active atomic electron.
This approach is well justified for negative ions, but
needs to be modified for the processes with neutral atoms.
Section \ref{eik} exposes our recent progress based on
the eikonal approach which allows us to take into account
the Coulomb field.
The concluding Section \ref{con} summarizes the results.
The atomic units are used
throughout the paper unless indicated otherwise.

\section{Factorization technique}
\label{fact}

The factorization technique of Ref.~\cite{K95}
can be applied to a number of multiphoton processes.
In this section our attention is restricted to an important
example of HG which has been recently considered by the
present authors \cite{KO98let,KO99pap}.
We concentrate on the linearly polarized laser field 
\begin{equation}\label{f(t)}
{\bf F}(t) = {\bf F} \cos \omega t 
\end{equation} 
that creates the external potential $V(t) = {\bf F}(t)\cdot {\bf r}$
acting on
atomic electrons.
Using the second order time-dependent perturbation theory
one can present the amplitude of HHG $d_N^+$ in 
the single-active-electron approximation in the following form
\begin{eqnarray}\label{dN}
d_N^+ = 
- \, \frac{i}{T}\int_0^T dt \,\int_{- \infty}^t dt^\prime \,
\langle \phi_a (t) \,|\, \exp (i \Omega t) \,
U \,G(t,t^\prime) \, V(t^\prime)\, | \,\phi_a (t^\prime) \rangle ~.
\end{eqnarray}
Here the brackets $\langle\, |\,|\, \rangle$ imply
integration over 
coordinates,
$T=2\pi/\omega $ is the laser period.
The initial-state wave function is 
\begin{equation}\label{phi}
\phi_a ({\bf r},t) = \phi_a ({\bf r}) \exp (-i E_a t)
\end{equation}
where $E_a= -\kappa^2/2$ is the bound state energy and
$\phi_a({\bf r})$ is the corresponding atomic eigenfunction.
The potential $U = \mbox{\boldmath $\epsilon$} \cdot {\bf r}$
accounts for
production of
the high harmonic with the frequency
$\Omega= N \omega$ and the polarization $\mbox{\boldmath $\epsilon$}$. 
The Green function $G(t,t^\prime)$ describes
the electron propagation in the intermediate state.
The amplitude (\ref{dN}) takes into account only 
a sequence of events in which HHG follows the absorption of the necessary
energy from the laser field. 
A number of omitted, the
so called time-reversed sequences,
is strongly suppressed due
to multiphoton nature of the process.

Neglecting the potential of the atomic core (a possible way 
to lift this approximation is considered in Section \ref{eik}), 
one can present the Green function via a complete set of 
the Volkov wave functions $\phi_{\bf p}({\bf r},t)$
that account for the electron dressing by the laser field
\begin{eqnarray}\label{G}
&&G({\bf r},t; {\bf r}^\prime, t^\prime)= \int\, \phi_{\bf p}({\bf r},t)
\phi_{\bf p}^*({\bf r}^\prime, t^\prime )\, \frac{d^3 p}{(2\pi)^3}~, 
\\ \label{V}
&&\phi_{\bf p}({\bf r},t) = 
\exp \left\{i \left[ \left( {\bf p}+\frac{\bf F}{\omega}
\sin \omega t \right)
\cdot {\bf r} -
 \, \frac{1}{2} \int^t \left( {\bf p}+\frac{\bf F}{\omega} 
\sin \omega \tau \right)^2 d \tau \right] \right\}~.
\end{eqnarray}
The multiphoton nature of
the problem makes the phases of the integrand in (\ref{dN})
to vary rapidly with $t$, $t^\prime$. This circumstance allows one to use 
the saddle-point approximation for integrations over
the time variables. After the {\it accurate}\/ integration 
over the momenta ${\bf p}$ in (\ref{G}), see details in 
\cite{K95,KO98let,KO99pap}, 
it is possible to show that (\ref{dN}), (\ref{G}), (\ref{V}) 
lead to the following convenient presentation of the HHG amplitude
\begin{eqnarray} \label{sumc} 
d^+_N & = & 2 \, \sum_m \, d^+_{Nm} ~,
\\ \label{dc}
d^+_{Nm} & = & A_{m \, \mu_0}({\bf K}_m) 
\, B_{N \, m \mu_0}({\bf K}_m) ~.
\end{eqnarray}
Each term in (\ref{sumc})
is written as a product of two
amplitudes of {\it physical, fully accomplished and observable}\/
processes; no "off-shell" entities appear.
Therefore Eqs.~(\ref{sumc}), (\ref{dc}) clearly demonstrate
the stepwise character of the process.
The {\it first step}\/ is described by the first factor 
$ A_{m \, \mu_0}({\bf K}_m)$ which is an amplitude of physical 
ATI process when after absorption of $m$ laser photons
the active electron acquires a translational momentum 
${\bf p} = {\bf K}_m$. 
In the Keldysh-type approximation 
this amplitude can be evaluated using the saddle-point technique
discussed in Section \ref{ele}.
The subscript $\mu_0$ in $A_{m \, \mu_0}$ specifies the contribution
of one particular saddle point in $t^\prime$ integration,
see more details in Section \ref{ele}.
The other factor, $B_{N \, m \mu_0}({\bf K}_m)$, is a combined
amplitude of the {\it second and third steps}, which are the
propagation and laser assisted recombination (PLAR). 
It can be further factorised into the propagation factor 
$1/R_{m \mu_0}$ 
describing the {\it second step}\/ and the amplitude of the 
{\it third step}, which is the laser assisted recombination (LAR),
$C_{N m}({\bf K}_m)$:
\begin{eqnarray} \label{prfact}
B_{N \, m \mu_0}({\bf K}_m)  \simeq \frac{1}{R_{m \mu_0}} \, 
C_{N m}({\bf K}_m) ~.
\end{eqnarray}
$R_{m \mu_0}$ is merely an approximate expression for the distance
passed by the active electron in course of its laser-induced
wiggling motion
\begin{eqnarray} \label{propap}
\frac{1}{R_{m \, \mu_0}} = - \, \frac{\omega^2}{F
\cos \omega t^\prime_{m \mu_0} } ~.
\end{eqnarray} 
The amplitude $C_{N m}({\bf p})$ of the physical LAR
process describes the laser assisted recombination, i.e.
transition of an electron with momentum ${\bf p}$ from 
the continuum to the bound state. Since the continuum state is 
laser-dressed, the recombining electron can emit the $N$-th 
harmonic photon, gaining necessary extra energy from 
the laser field. More details on LAR amplitude
are given in Section \ref{ele}.

The summation in formula (\ref{sumc}) runs over a number of photons 
$m$ absorbed on the first step of the process when the active 
electron is released. 
In other words one can say that the energy conservation constraint
selects the discrete set of ATI channels in the laser-dressed 
continuum. These channels serve as intermediate states for 
the three-step HG process. 
In a given channel the electron has translational momentum ${\bf K}_m$
with the {\it absolute value}\/ defined by
\begin{equation}\label{Kmnm}
K_{m} =\sqrt{ 2 \left( m \omega - U_p
+ E_a \right) } ~.
\end{equation}
Here $U_p \equiv F^2/(4 \omega^2)$ is 
the well-known ponderomotive potential.
ATI plays a role of the first stage of HHG process only if
the electron momentum has specific {\it direction},
namely ${\bf K}_m$ is directed along ${\bf F}$.
This ensures eventual electron return to the core that makes
the final step, LAR, possible as discussed in detail in
Refs.~\cite{K95,KO98let,KO99pap}.

The observable HG rates ${\cal R}_N$ are expressed via
the amplitudes as 
\begin{eqnarray} \label{RN}
{\cal R}_N \equiv 
\frac{\Omega^3}{2 \pi c^3} \, 
\left| d_N^+ \right|^2 ~,
\end{eqnarray}
$\Omega=N\omega$ is the frequency of emitted harmonic, 
$c$ is the velocity of light.

Eq.~(\ref{sumc}) is the major result of this section.
It presents the amplitude of the ``complicated'' HG process
in terms of the amplitudes of ``elementary'' processes which 
are the ionization and LAR. 
Conceptual significance of this result is based on the fact that it supports 
the three-step interpretation of the HG process
which, in turn, stems from the physical picture of the
atomic antenna discussed in Section~\ref{intro}.
Eq.~(\ref{sumc}) is also very convenient for numerical applications, since 
a knowledge on sufficiently simple elementary amplitudes enables one to
calculate accurately the HG process, as discussed below in Section~\ref{hhg}.

\section{Effective channels}
\label{eff}

The summation over $m$ in Eq.~(\ref{sumc}) has a clear physical
interpretation. After the first step (ionization) the released
electron can  be found in any ATI channel before colliding
with the parent atom. After the third step all intermediate
channels result in the same final electron state.
Therefore the contributions of intermediate ATI level interfere,
as shows the summation over $m$ in Eq.~(\ref{sumc}). 
The interference has several prominent manifestations including
the cutoff of the HHG rates for high $N$ and their
oscillations in the plateau domain,
as discussed in Section~\ref{hhg}.
The number of the ATI channels which give significant contribution
to the amplitude is usually large 
\begin{equation}\label{dm}
\delta m \gg 1~.
\end{equation}
This circumstance does not 
pose a problem for numerical applications based on Eq.~(\ref{sumc}), but
may in some situations obscure the qualitative analyses. 
To overcome this difficulty
it is desirable to carry out
summation over $m$ in (\ref{sumc}) in an analytical form.
This is the major task of this Section which follows the approach of 
Ref.~\cite{KO00eff}.

Let us verify first that the summation over $m$ in
(\ref{sumc}) can be replaced by integration over the related
continuum variable
\begin{equation}\label{int}
d^+_N  = \sum_m  d^+_{Nm}  \simeq \int \, d^+_{Nm}\, dm ~.
\end{equation}
To prove the reliability of the approximation based on
(\ref{int}) we use the Poisson summation formula
which allows one to present the amplitude of HHG in the following form
\begin{equation}\label{pi}
\sum_m d^+_{Nm}  = \sum_j \int  dm \,\exp (- 2\pi i j m)\, d^+_{Nm} ~.
\end{equation}
The summation index  $j$ in
(\ref{pi}) may be looked at as a variable which is
conjugate to the $m$ variable. Since $m \omega$ 
refers to the spectrum,
$2 \pi j/\omega$ should be identified as a
time variable or, more accurately, as
a number of periods of
time that elapsed between the ejection of the electron from an atom
and its return back. The wave function of the released
electron spreads in space the more the larger $j$ is. Therefore
one should anticipate that the most important contribution
to (\ref{pi}) is given by the major term $j=0$. 
Same conclusion can be drawn from Eq.~(\ref{int}). 
The estimate shows that the spectral
variable $m$ covers a wide region
which causes the time variable $j$ to be localized. 
This discussion demonstrates that one can safely take
the leading $j=0$ term in (\ref{pi}) thus supporting (\ref{int}).

An advantage of integration over the variable $m$ in (\ref{int})
is that it can be carried out
in closed analytical form using the saddle-point method.
To specify this statement
let us return to
Eqs.~(\ref{dN}), (\ref{G}), (\ref{V}). One can deduce from them that 
the major dependence of the phase
of the integrand of (\ref{dN}) on integration variables is 
associated with the factor $\exp (i{\cal  S})$ where ${\cal S}$
is the classical action
\begin{eqnarray} \label{genS}
{\cal S} = {\cal S} (t, t^\prime, m) = 
\int^{t}_{t^\prime} 
d \tau \left[\frac{1}{2} \left({\bf K}_m + 
\frac{{\bf F}}{\omega} \sin \omega \tau \right)^2
-E_a \right] + \Omega t ~.
\end{eqnarray}
Since we know from Eq.~(\ref{dc}) 
that the momentum variable arises 
in the final formulas as ${\bf K}_m$, 
we can use this momentum in (\ref{genS})
instead of the integration variable ${\bf p}$
that originates from (\ref{G}). It should be noted  
that Eq.~(\ref{dc}) where ${\bf K}_m$ arises
was derived using {\it accurate}\/ integration
over {\it all momenta}\/ ${\bf p}$; hence the substitution
${\bf p}\rightarrow {\bf K}_m$ in (\ref{genS})
is not an additional approximation.

Eq.~(\ref{sumc}) was obtained using the saddle points approximation
for integrations over the time variables $t$, $t^\prime$. 
Positions of $m$-dependent saddle points $t_m$, $t^\prime_m$
are governed by the equations
\begin{eqnarray}\label{t}
\frac{ \partial}{\partial t} \, {\cal S} (t, t^\prime, m) & = & 0 ~,
\\ \label{t'}
\frac{ \partial}{\partial t^\prime} \, {\cal S} (t, t^\prime, m) & = & 0 ~,
\end{eqnarray}
in which $m$ is considered as an integer labeling
the physical ATI channel.
As was mentioned above, Eq.~(\ref{int}) opens a possibility for
integration over $m$ that can also be carried out
using the saddle point approximation. 
The positions of corresponding saddle
points are governed by the following equation
\begin{eqnarray}\label{m}
\frac{\partial}{\partial m} \, {\cal S} (t, t^\prime, m) = 0 ~,
\end{eqnarray}
where one can write the partial derivative over $m$ because
the $m$-dependence of $t_m$ and $t^\prime_m$ does not contribute
due to (\ref{t}), (\ref{t'}). Eqs.~(\ref{t'}), (\ref{t}), (\ref{m})
define two instants of time $t_m$, $t_m^\prime$ at which, respectively,
the electron emerges from an atom and returns back to it,
as well as the number of quanta $m = m_{\rm eff}$ absorbed 
in course of the ionization.
All these three variables are, generally speaking, 
the complex-valued functions of the frequency 
$\Omega = N\omega$ 
of the generated harmonic. For a given $N$ there can be several 
solutions of Eqs.~(\ref{t}), (\ref{t'}), (\ref{m}).

Integrating over $m$ in Eq.~(\ref{int}) by the saddle method 
gives the following representation for the HHG amplitude
\begin{equation}\label{s''}
d^+_N =  2 \, \sum_{m_{\rm eff} } \, 
\left( \frac{2\pi}{ i S^{\prime\prime}_{m_{\rm eff}} } \right)^{1/2} \,
d^+_{Nm_{\rm eff}} ~,
\end{equation}
which is the major result of this Section.
Comparing (\ref{s''}) with (\ref{sumc}) one finds,
along with clear similarities, several distinctions.
The most substantial of them originates from different 
physical meaning of the summation index in these formulas.
In Eq.~(\ref{sumc}) it is an integer labeling channels
in the {\it physical}\/ ATI spectrum.
In contrast, in formula (\ref{s''}) summation runs over
complex-valued $m_{\rm eff}$ set. It is natural to say
that $m_{\rm eff}$ are labels of {\it effective channels}.
In order to find the amplitude $d^+_{Nm_{\rm eff}}$ one can use 
representation (\ref{dc}) for $d^+_{Nm}$
and continue the amplitudes 
$A_{m \, \mu_0}({\bf K}_m)$ and $B_{N \, m \mu_0}({\bf K}_m)$ 
into the complex-$m$ plane
which can be done if they are known sufficiently well.
One more, though less important difference, is presented by
an additional square root factor in (\ref{s''})
that arises from integration over $m$ and depends on
the second derivative $S^{\prime\prime}_m$ of the action
(\ref{genS}) over $m$.

The advantage of effective channel representation stems from
the fact that only small number of effective channels
(actually one or two) contributes, whereas the number
of essential real ATI channels is quite large, as discussed
above. Bearing this in mind it is worthwhile to illustrate
variation of effective channels labels $m_{\rm eff}$
with $N$ by solving numerically set of equations (\ref{t'}),
(\ref{t}), (\ref{m}) for some particular case \cite{KO00eff}. 
Fig.~1 shows two important solutions $m_{\rm eff}(N)$ that
move along trajectories in the complex-$m$ plane as $N$ varies.
The overall picture comprises a characteristic cross-like pattern. 
For small $N$ the trajectories are close to the real axis
and are well separated. They approach each other as $N$ increases
and almost "collide" at some particular critical value $N=N_c$.
For larger $N$, $N>N_c$, the trajectories start to move
almost perpendicular to the real-$m$ axis and rapidly
acquire large imaginary parts.
\begin{figure}[t]
\label{comp}
\mbox{ \hspace{3.7cm}\psfig{file=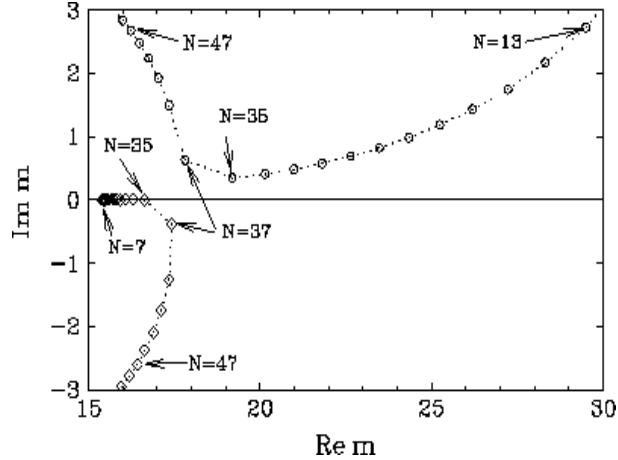,width=8cm,height=6cm}}
\vspace{-2mm} 
\caption{\small{
Trajectories of the effective channel labels $m_{\rm eff}(N)$
in the complex $m$-plane for laser frequency $\omega = 0.0043$a.u.,
intensity $I$ = 10$^{11}$ W/cm$^2$ and varying
harmonic order $N$ for HHG by H$^-$ ion.
Positions of two effective channel labels for odd integer $N$
are denoted respectively by circles and diamonds.
} }
\end{figure}
It can be demonstrated that large ${\rm Im} (m_{\rm eff})$
lead to suppression of the HHG process.
Therefore the critical $N_c$ marks the beginning of 
the cutoff region for HHG. As shown in Ref.~\cite{KO00eff},
approximate analytical solution of
Eqs.~(\ref{t'}), (\ref{t}), (\ref{m})  
shows that the critical value $N_c$ is equal to
\begin{equation}
N_c \,\omega = |E_a| + 3.1731 \,U_p~,
\end{equation}
in agreement with the well known result of 
Refs.~\cite{KulanderA,KulanderB,Lew,BLM94}.

In order to find simple
physical interpretation for the effective channels, let us 
note that Eqs.~(\ref{t}), (\ref{t'}), (\ref{m}) may be considered as
classical equations of motion in the laser field. 
They define the classical trajectories along
which the electron first goes away from the atom,
and then returns back accumulating during this motion
energy from the laser field that is necessary for HG.
This physical picture agrees with the atomic antenna concept discussed
in Section~\ref{intro}. It also comes in line with the
Corcum model \cite{C93} based entirely on the classical trajectories.
Even more close relation can be found
with the approach of Lewenstein {\it et al}\/ \cite{Lew}, which uses the
saddle point method to integrate over the momenta ${\bf p}$
in formula (\ref{G}).

Eqs.~(\ref{sumc}) and (\ref{s''}) provide two
different ways to describe the atomic antenna concept, either
in terms of real physical channels
in the intermediate ATI spectrum, or 
by using the effective channels for the intermediate state. 
Each approach has its advantages which can be beneficial
for different aspects of HHG problem. 
Importantly, the above discussion 
ensures identity of the two formulas since 
(\ref{s''}) was derived directly from (\ref{sumc}).

\section{Photoionization and recombination}
\label{ele}

Eq.~(\ref{sumc}) shows that 
the process of HHG is intimately related to 
ionization and LAR.
This fact makes the latter processes
very interesting from the perspective of the atomic antenna concept,
in addition to their well known importance as the basic
events
in the laser-matter interaction.
This Section describes the recent progress in the theory of 
these two
"elementary"
phenomena.
\begin{figure}[h]
\begin{center}
\label{1-3} 
\raisebox{0cm}
{\mbox{ \hspace{0cm}\psfig{file=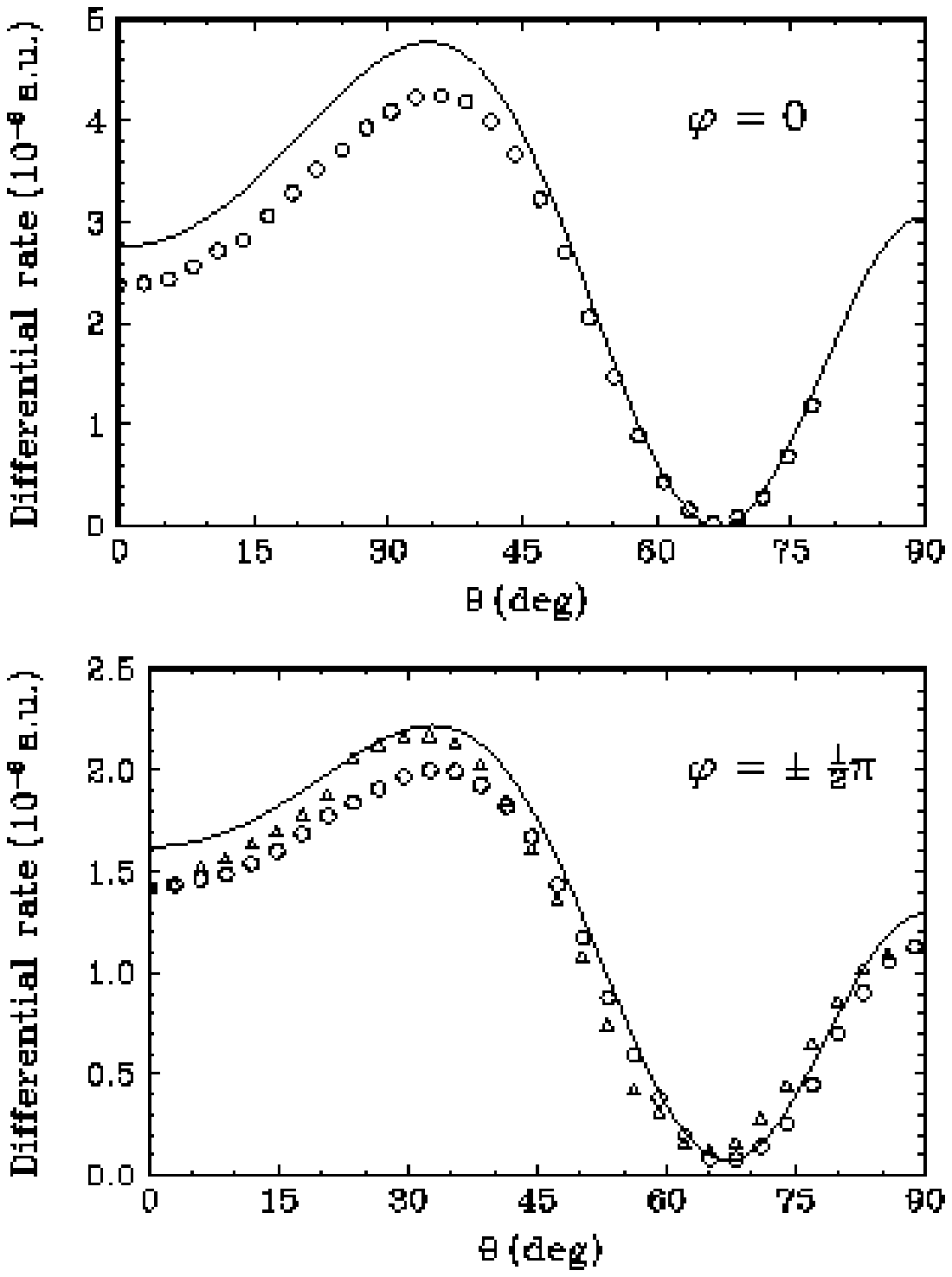,width=7.2cm,height=9cm}} }
\raisebox{-0.04cm}
{\mbox { \hspace{0.3cm}\psfig{file=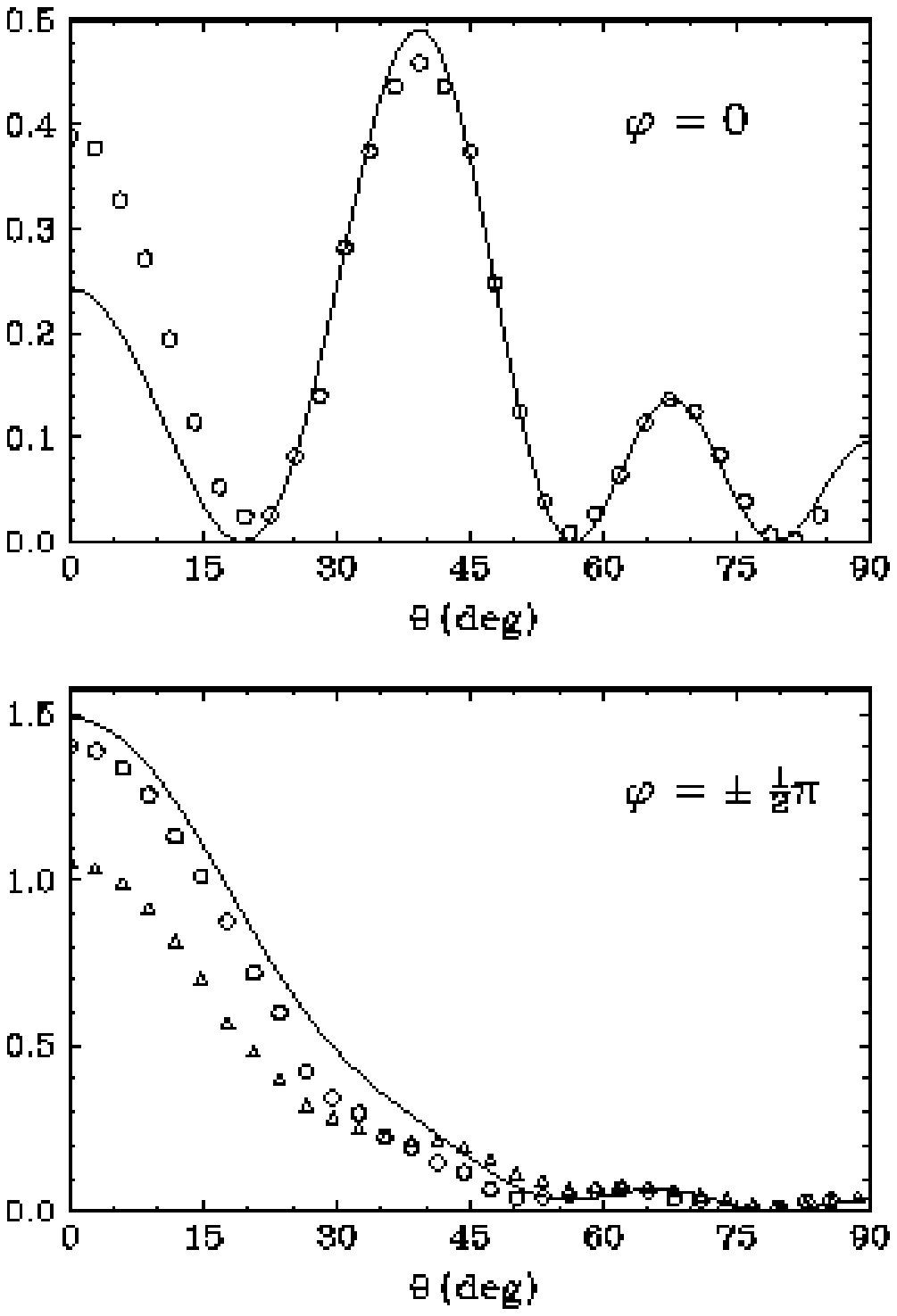,width=6.8cm,height=9.1cm}  }}
\vspace{-2mm} 
\caption{\small{
Detachment of H$^-$ ion in bichromatic field
with the frequencies $\omega = 0.0043$ a.u. and $3\omega$
and intensities $I_1 = 10^{10} {\rm W}/{\rm cm}^2$
and $I_2 = 10^{9} {\rm W}/{\rm cm}^2$ respectively.
Differential detachment rate 
(in units $10^{-8}$a.u.) as a function of the electron
emission angle $\theta$
is shown for  various values
of the field phase difference $\varphi$ as indicated in the plots.
Open symbols show results of calculations of \protect\cite{T} 
(in the $\varphi = \pm \frac{1}{2} \pi$ plot the open  
circles show the results for $\varphi = \frac{1}{2} \pi$ and
open triangles these for $\varphi = - \frac{1}{2} \pi$).
Solid curves show results of the  adiabatic theory \protect\cite{KOdetach99}
(which coincide for $\varphi = \frac{1}{2} \pi$ and
$\varphi = - \frac{1}{2} \pi$).
Left - the first ATD peak, corresponding to absorption 
of $n=8$ photons of frequency $\omega$.
Right -  the third ATD peak, corresponding to
absorption of $n=10$ photons of frequency $\omega$.}}
\end{center}
\end{figure}
Consider first the multiphoton ionization. 
Adopting the Keldysh-type approach which neglects the
field of the atomic core in the final state one can present 
the amplitude of the ionization in the following
form
\begin{equation}\label{kel}
A_m({\bf p}_m) = \frac{1}{T}\int_0^T \langle \phi_{ {\bf p}_m}(t)\, | 
\, V(t) \, | \, \phi_a(t) \rangle ~,
\end{equation}
where $\phi_{\bf p}(t)= \phi_{\bf p}({\bf r},t)$
is the Volkov wave function (\ref{V})
and ${\bf p}_m$ satisfies the energy conservation law
\begin{equation}\label{pen}
E_a+m\,\omega = \frac{1}{2} \, {\bf p}_m^2 + U_p ~.
\end{equation}
Fast variation of the phase of the integrand in
(\ref{kel}) allows one to use the saddle-point
approach for integration over the time $t$. This
approximation,
first proposed
by Keldysh \cite{Keldysh}, was developed in detail in Refs.~\cite{P1,P2,P3}.
Using this scheme one presents the photoionization amplitude as 
\begin{equation}\label{kel1}
A_m({\bf p}_m) = \sum_\mu A_{m\mu}({\bf p}_m)~, 
\end{equation}
where summation runs over essential saddle points labeled by
subscript $\mu$.
Note that in Eq.~(\ref{sumc}), relating the ionization
problem with HG, it is sufficient to take into account
the amplitude which arises from only one saddle point (of two)
labeled as $\mu_0$;
another saddle point gives the same contribution that produces
a factor 2.

The technique described above was refined in Refs.~\cite{GK97pra,GK97jpb}.
In the pioneer publications \cite{Keldysh,P1,P2,P3}
the momentum $p$ of the ionized electron was treated as a small
quantity, and all essential functions were expanded in powers
of $p/\kappa$. This approximation restricted the applicability 
of the approach, since for high channels
in ATI spectrum the momentum is not small.
Refs.~\cite{GK97pra,GK97jpb} demonstrated that
the technique can be modified to include large electron momenta as well.
Importantly, this modification retains
simplicity and clear physical nature of the Keldysh approach.

One can anticipate that the Keldysh-type methods
should produce reliable results for photodetachment
of
negative ions,
where the detached electron is
influenced
negligibly
by the Coulomb
field
of the residual atomic particle.
Due to this reason  the photodetachment 
was in the focus of attention of
Refs.~\cite{GK97pra,GK97jpb}
which compared the
results of
improved Keldysh approximation
with a variety of numerical and experimental data available
for negative ions.
Refs.~\cite{KOdetach98,KOdetach99} continued this study 
and extended it to the case of the two-color laser field.
Detailed description of all results obtained in these works
would bring us too far away from the main topic of this paper.
However, it is important to mention that the overall accuracy
of the modified Keldysh approximation proves be very high. 
It
closely
reproduces results of other, much more sophisticated
methods for total probabilities of detachment as well as for
spectral and angular distributions of photoelectrons both for
the weak and strong field
regimes
(i.e., for any
value of the Keldysh parameter $\gamma = \kappa \omega/F$).
An example of photodetachment of the H$^-$ in bichromatic laser field
with the frequencies $\omega = 0.0043$ a.u. and $3\omega$
shown in 
Fig.~2 illustrates this point.

Let us now turn our attention to the other
relevant problem, laser-assisted 
photo recombination (LAR). Consider the electron-atom
impact in a laser field which results
in creation of a negative ion and HG. Since the system can
acquire energy from the laser field due to absorption of several
laser quanta, the emitted harmonics should exhibit the 
equidistant spectral distribution. 
Strange enough, this important process was not studied theoretically
until two recent almost simultaneous publications \cite{KO00,Jar}.

The amplitude of the photorecombination
can be written as
\begin{equation}\label{rec}
C_m({\bf p}) = \frac{1}{T}\int_0^T \langle \phi_a(t) \, | 
\exp (i\Omega t)\, U \, | \phi_{\bf p}(t) \rangle~.
\end{equation}
Here  $U = \mbox{\boldmath $\epsilon$} \cdot {\bf r}$
describes the potential which is responsible for the harmonic production,
and $\Omega$ is the energy of the generated harmonic which satisfies
the energy conservation constraint
\begin{equation}\label{recec}
\Omega = \frac{{\bf p}^2}{2} + U_p + | E_a| + m\,\omega~,
\end{equation}
in which $m$ is the number of laser photons absorbed
during recombination.
The wave function of the electron in the continuum in Eq.~(\ref{rec}) 
can be described by the Volkov wave function, similarly to
the Keldysh-type approach to photoionization.
Comparing  the amplitudes 
(\ref{rec}) and (\ref{kel}) one observes their close similarity.
This fact allows one to develop 
the theory on the basis of formula (\ref{rec})
along the lines described above for the ionization
problem. In particular, one can use the saddle-point approximation
for integration over the time variable in (\ref{rec}).

This approach, suggested in Ref.~\cite{KO00}, is supplemented
in the cited paper by several numerical examples. One of them,
shown in Fig.~3,  depicts the cross section of recombination 
on hydrogen atom in a laser field with
$\omega = 0.0043$ a.u. and the intensity $I = 10^{11}$ W/cm$^2$
versus the energy of the emitted high harmonic.
\begin{figure}
\label{lar}
\mbox{ \hspace{2.5cm}\psfig{file=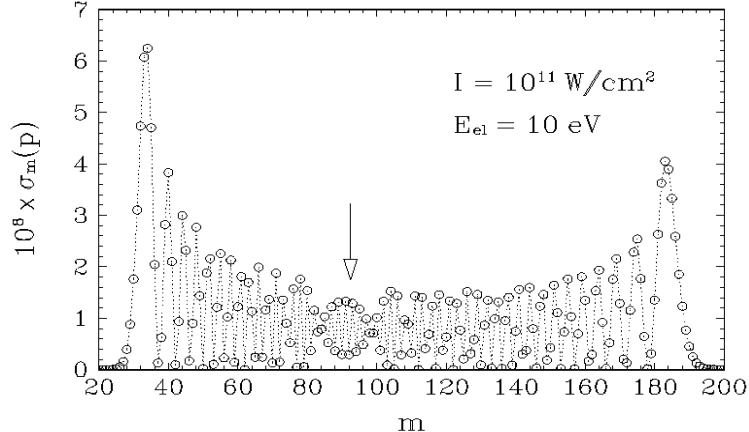,width=10cm,height=6cm}}
\caption{ \small{ 
Cross section 
$\sigma_m(p)$ 
for laser-assisted recombination of the electron 
with the energy $E_{\rm el} = 10$ eV into the bound state in H$^-$ ion
\protect\cite{KO00}.
The results  are shown for the laser field with the frequency 
$\omega = 0.0043$ a.u.  and the intensity $I = 10^{11}$ W/cm$^2$. 
The symbols are joined by lines to help the eye.
}}
\end{figure}
Since this is the first work in the field we 
could not compare our results with
other calculations. Bearing in mind that
the recombination process has similarities with
the ionization problem, where similar approach
works well
one can expect reliable results in the
recombination problem as well.

\section{
Quantitative illustrations for HHG}
\label{hhg}

To illustrate the applicability of the two methods, 
the factorization technique and the method of effective channels, 
consider an example of HHG
by
hydrogen negative ion
in a laser field with $\omega = 0.0043$ a.u.$\,$.
Using the factorization procedure
one needs first to calculate the amplitudes
of ATI and LAR, which can be done
by
the technique
discussed in Sections \ref{ele}. After that 
employing
(\ref{sumc}) one finds the amplitude of HHG, and
from (\ref{RN}) the HHG rates. The results are presented in Fig.~4.

It is important to note
a strong interference of contributions coming from
different intermediate channels $m$. 
In the plateau region 
it is responsible for an oscillatory pattern, and
becomes even  more important 
in the cutoff region, where a contribution of
any
single channel
drastically exceeds  the results of an accurate 
summation over a large number of channels.
\begin{figure}[h]
 \begin{center}
\mbox{
\hspace{-0.3cm}\psfig{file=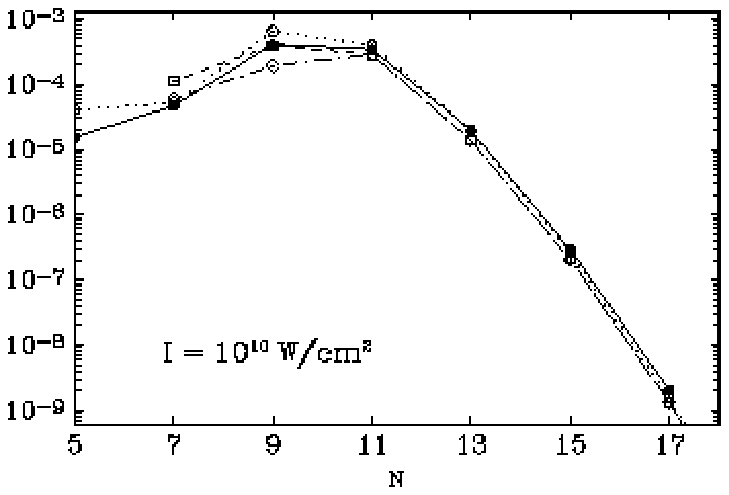,width=7.7cm,height=5.2cm} 
\raisebox{0.1cm}{\mbox { \hspace{0cm}\psfig{file=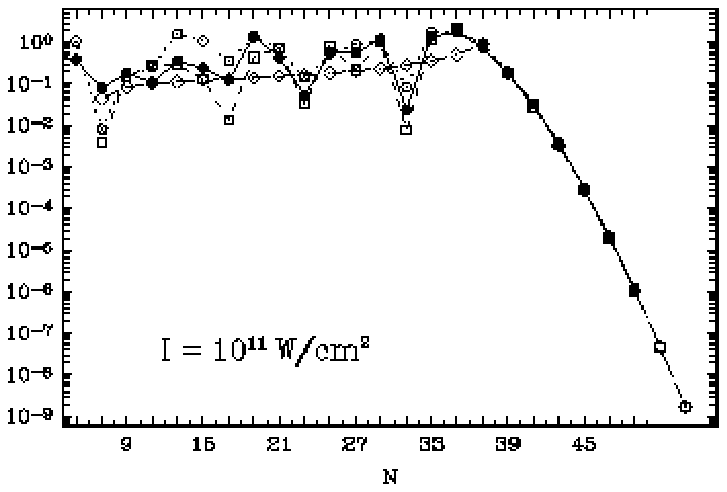,width=7.7cm,height=5.08cm}  }}} 
 \end{center}
\label{hhgr}
\vspace{-2mm} 
 \caption{ \small{ 
Harmonic generation rates 
(\protect \ref{RN})
(in sec$^{-1}$) for H$^-$ ion 
in the laser field with the frequency $\omega = 0.0043$ a.u. 
and various values of intensity $I$ as indicated in the plots.
Closed circles - results obtained by Becker {\it et al}\/
\protect\cite{BLM94}, open circles - calculations
\protect\cite{KO98let,KO99pap}
based on factorization procedure and
performing numerical summation (\protect\ref{sumc}) over 
contributions of different ATI channels, open diamonds -- 
calculations of Ref.~\protect\cite{KO00} based on the 
effective channels approach (\protect \ref{s''})
with a single effective channel $m_{\rm eff}(N)$
taken into account (namely, the effective channels shown by 
diamonds in Fig.~1); 
open squares -- same but taking into account two saddle
points $m_c(N)$ (namely, the effective channels shown by 
diamonds and circles in Fig.~1).
The symbols are joined by lines to help the eye.}}
 \end{figure}
In order to apply the technique based on 
the effective channels one needs to
calculate the "elementary" amplitudes for the complex-valued
number $m_{\rm eff}$ of quanta
absorbed on the first step of three-step process.
This can be done by the
approach described in Section \ref{eff}, because it
relies on analytical methods for calculation of these amplitudes
that remain valid for a complex-valued $m$.
Taking the effective channels (presented in Fig.~1 for 
$I$ = 10$^{11}$ W/cm$^2$),
calculating for them the "elementary" amplitudes and
applying formula (\ref{s''}), we find the rates presented in Fig.~4.
If one takes into account in 
the summation
(\ref{s''})
over $m_{\rm eff}$
a single effective channel, shown by diamonds in Fig.~1,
then the cutoff region is nicely
described, as well as the overall pattern in the plateau 
domain. However such one-saddle-point approximation does 
not reproduce structures in the $N$-dependence of HG rates. 
Taking into account the two effective channels,
shown by diamonds and circles in Fig.~1
improves the results for 
the rates in the plateau domain by producing appropriate structures 
in the $N$-dependence. Remarkably, this two-saddle-point
calculation gives correct positions of minima and maxima
in the rate $N$-dependence, albeit the magnitudes of
the rate variation is reproduced somewhat worse; for
instance the depth of the minimum at $N=17$ is quite
strongly overestimated. 

Fig.~4 shows good agreement of the approaches based on the
factorization technique and on the effective channels, 
which both
are in accord
with the results of Ref.~\cite{BLM94}.
This agreement holds both for multiphoton regime
(left part of Fig.~4), as well as 
in the tunneling regime (right part of Fig.~4).

\section{Above-Threshold Ionization
in high channels
}
\label{ati}
The methods discussed in Sections \ref{fact}, \ref{eff}
can be applied to a number of other multiphoton problems. 
To illustrate this point consider the factorization technique
for ATI. The ionization amplitude  $A_n({\bf p}_n)$ in
the Keldysh-type approximation, considered in (\ref{kel}),
neglects interaction of the released electron with  the core.
Let
$A^{(1)}_n({\bf p}_n)$
be
a correction that
takes this interaction into account.
In this notation the total amplitude
of ATI with absorption of $n$ photons is 
$A^{\rm tot}_n({\bf p}_n)=  A_n({\bf p}_n)+ A^{(1)}_n({\bf p}_n)$.
Using the approach of Ref.~\cite{K95}
we find  the following relation 
\begin{equation}\label{A0A1}
A^{(1)}_n({\bf p}_n) 
= \sum_m \sum_\mu  \sum_{\sigma=\pm1} 
A_m(\sigma {\bf K_m}) \frac{1}{R_{m\mu}}
f^B({\bf p}_n,\sigma {\bf K}_m)~,
\end{equation}
which presents the sufficiently complicated correction
$A^{(1)}_n({\bf p}_n)$ in terms
of two "elementary" amplitudes $A_m({\bf K}_m)$ and 
$f^B({\bf p},{\bf K})$. The later one describes
the electron-atom impact in the laser field in
the Born approximation.
This scattering can be named quasielastic, since the 
atom remains in the same state, but the electron momentum
is changed ${\bf K}\rightarrow {\bf p}$ both in direction
and absolute value.
Summation over $\mu$ in formula (\ref{A0A1})
reflects the fact that the electron emission into 
the continuum takes place at two moments of
time $t_{m\mu}$ ($\mu =1, \, 2$) per laser period.  
The electron return to the atom
is ensured only if
the electron
momentum is parallel (for one value of $\mu$) or
antiparallel (for another $\mu$)
to the field, see details in Ref.~\cite{K95}. 
This fact is taken
into account
by a summation index $\sigma=\pm 1$ in (\ref{A0A1}).

Eq.~(\ref{A0A1}) is obtained via application of the factorization
technique to ATI. It has a transparent physical meaning.
Ionization with absorption of $n$ photons
needs that first $m$ quanta are absorbed by an atom
removing the electron from an atom into the continuum state
with momentum $\pm{\bf K}_m$. 
The collision of this electron with
the atom
(often referred to as {\it rescattering})
results in absorption of additional $n-m$ quanta
and transition of the electron into a
state with momentum ${\bf p}_n$.
All intermediate ATI channels labeled by index $m$ contribute
coherently.
This physical picture of ATI agrees with the atomic antenna concept, 
as was first discussed in \cite{Ku87}.

We applied (\ref{A0A1}) to calculation 
of Above Threshold Detachment (ATD) from H$^-$ ion.
Fig.~5  shows the results which clearly indicate that
the contribution of the process (\ref{A0A1})
to the angular distributions of ATD spectra is
dominating
for higher ATD channels
while for low channels rescattering effects are small
.
\begin{figure}
 \begin{center}
\mbox{
\hspace{-0.5cm}\psfig{file=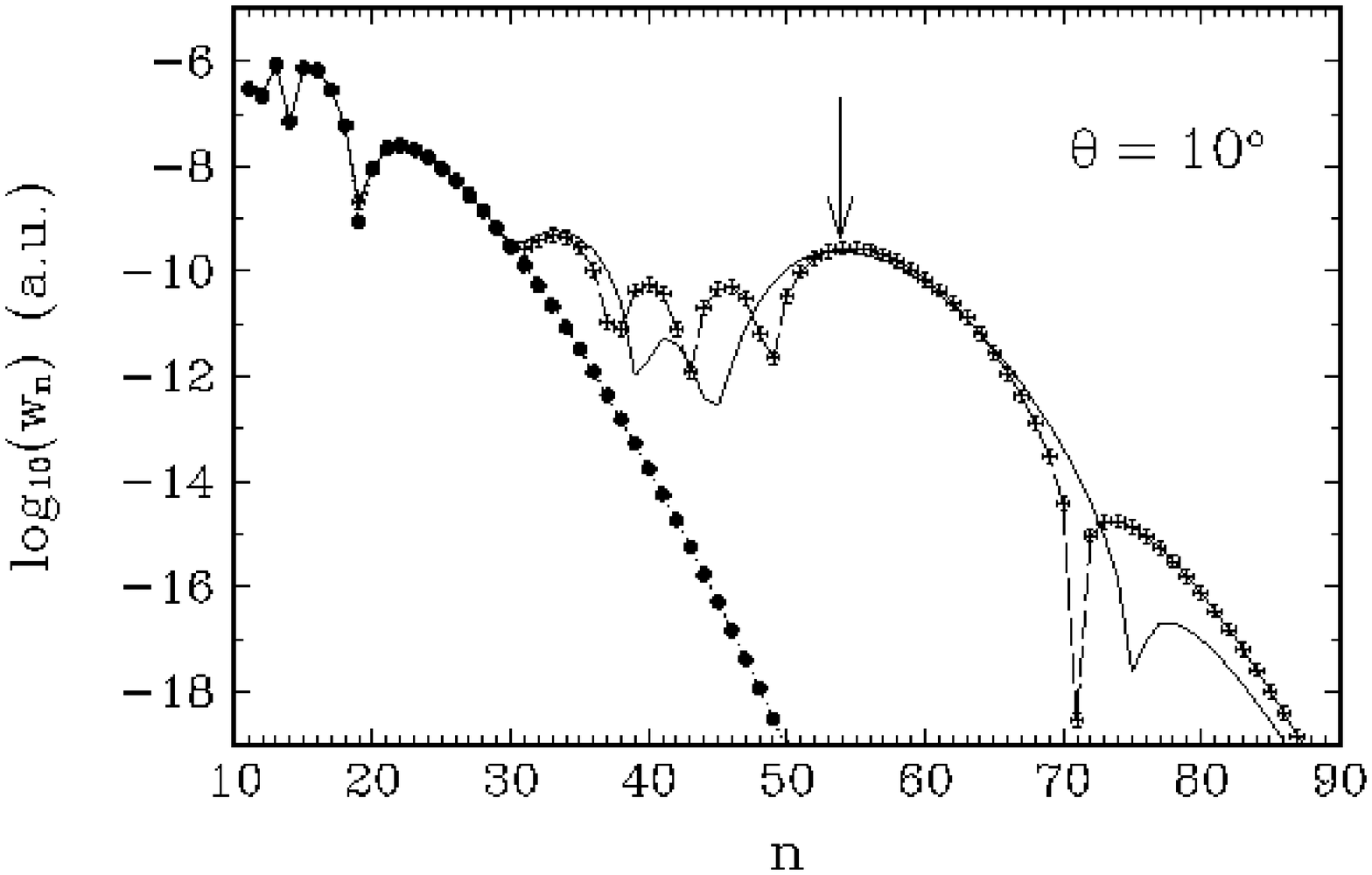,width=7.7cm,height=5.2cm} 
\raisebox{-0.05cm}
{\mbox { \hspace{0cm}\psfig{file=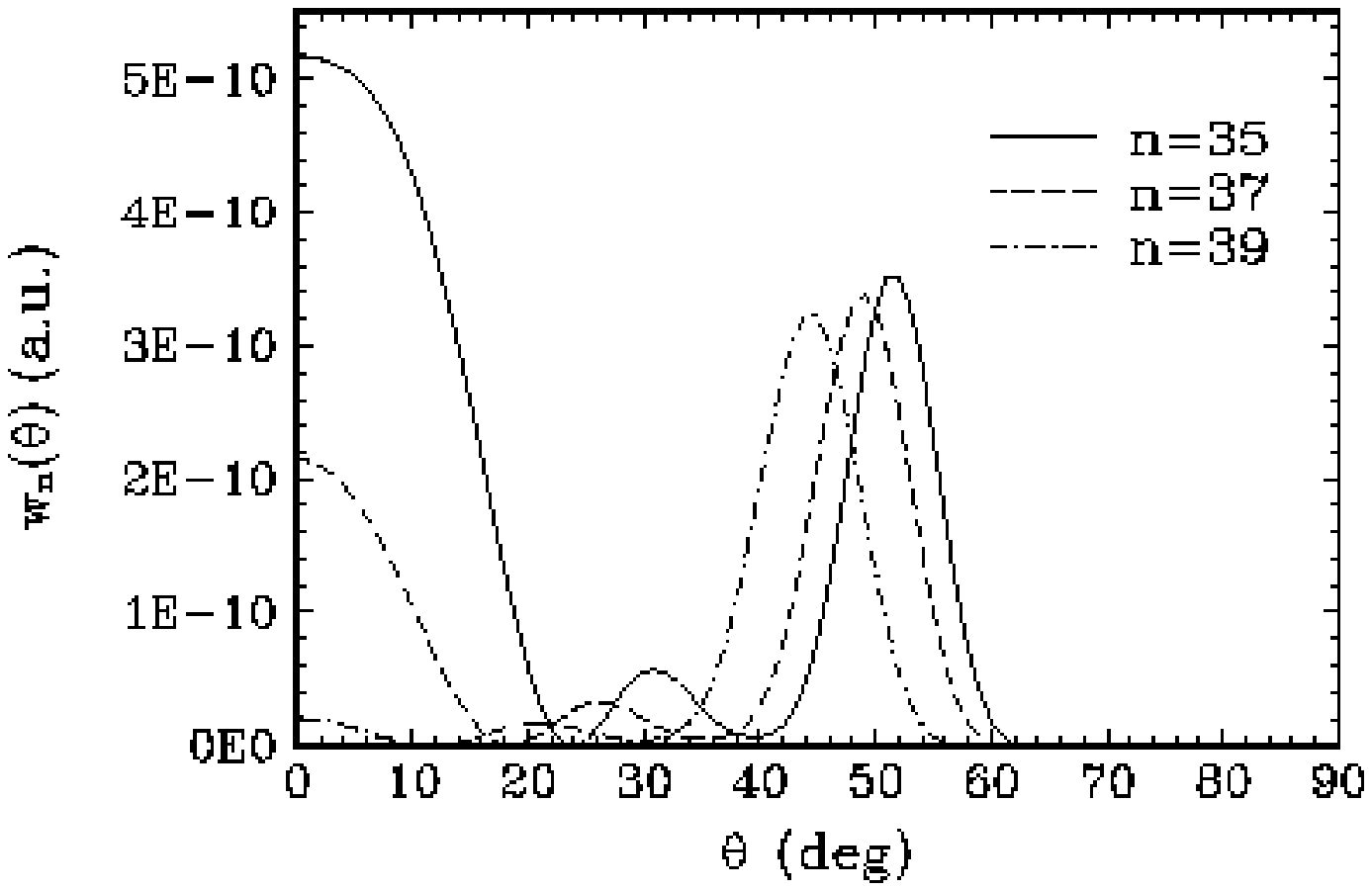,width=7.9cm,height=5.25cm}  }}
} 
 \end{center}
\label{atir}
\vspace{-2mm} 
 \caption{ \small
Differential above-threshold detachment rates $w_n(\theta)$. 
Left - the rates versus the number $n$ of absorbed photons
for fixed value of electron ejection angle $\theta$.
Right - same as functions of $\theta$ for fixed $n$.
An H$^-$ ion is irradiated by the laser wave with frequency
$\omega = 0.0043$ a.u. and intensity $I= 5\times 10^{10}$W/cm$^2$.
The minimum number of photons necessary for
ionization
is
$n_{\rm min}=11$. Curves -
present calculation with account of rescattering;
closed circles - Keldysh-type approximation
of \protect\cite{GK97pra}.
The arrow indicates the plateau cutoff as predicted by classical
theory \protect\cite{Pau}.}
\end{figure}

\section{Eikonal approach to the Coulomb field}
\label{eik}

The Keldysh-type approximation for laser induced ionization,
as described in Section \ref{ele}, discards interaction
of receding photoelectrons with the atomic core.
The most significant part of this interaction arises due to
the Coulomb field  of the core.
That is why a number of numerical applications
considered above
is carried out
for negative ions, where
the core is neutral.
However, for the multiphoton processes with neutral atoms, the Coulomb
forces between the active electron and the residual positive ion 
become operative. This interaction was taken into account first 
in Ref.~\cite{P1} in the quasistatic limit of the small Keldysh
parameter $\gamma = \kappa \omega/F \ll 1$. A very simple 
relation was obtained between
the photoionization rate $w_C$ for the electron bound by 
a potential with the Coulomb field produced by a core charge $Z$
and its counterpart $w_{\rm sr}$ for the electron 
with the same binding energy $\kappa^2/2$ but
bound by short range forces:
\begin{equation}\label{coul}
w_C = \left( \frac{2 \kappa^2}{F}\right)^{2Z/\kappa}w_{\rm sr}~.
\end{equation}
This result means that for conventional conditions the
presence of the Coulomb field enhances the 
rates by several orders of magnitude. 
It is remarkable that the relation (\ref{coul}) holds
in fact for arbitrary value of the Keldysh parameter, both
in the multiphoton and tunneling region, as was 
established in a more elaborate theory by Perelomov and Popov \cite{P3}.
This result agrees well with experimental data
for the total rates \cite{Chin}.

The theory of Perelomov and Popov is restricted to ejection of
low-energy electrons. It is usually anticipated that
these electrons give principal contribution
to the total rates summed over all ATI channels as well as over
ejection angles. The current experiments, however, are able to select an
individual ATI channel even for a large number of absorbed quanta.
Both energy and angular distributions of these electrons manifest
some fascinating features which are the object of interest in
modern experiment and theory. This fact prompts to
develop a theory which, reproducing the Perelomov-Popov results
for low electron energy, could also describe
high-energy
photoelectrons.
The antenna-type phenomena considered in this paper
provide an additional inspiration for this study.
The ionization amplitude in 
Eqs.~(\ref{sumc}), (\ref{A0A1}) should be summed over the
number $m$ of quanta absorbed that should be
large enough. This makes the electron momentum in the
intermediate ATI channels to be also large.
Therefore we need to know how the Coulomb field affects
the amplitudes for large momenta.
\begin{figure}[h]
\label{eiko}
\begin{center}
\mbox{ \hspace{-0cm} \psfig{file=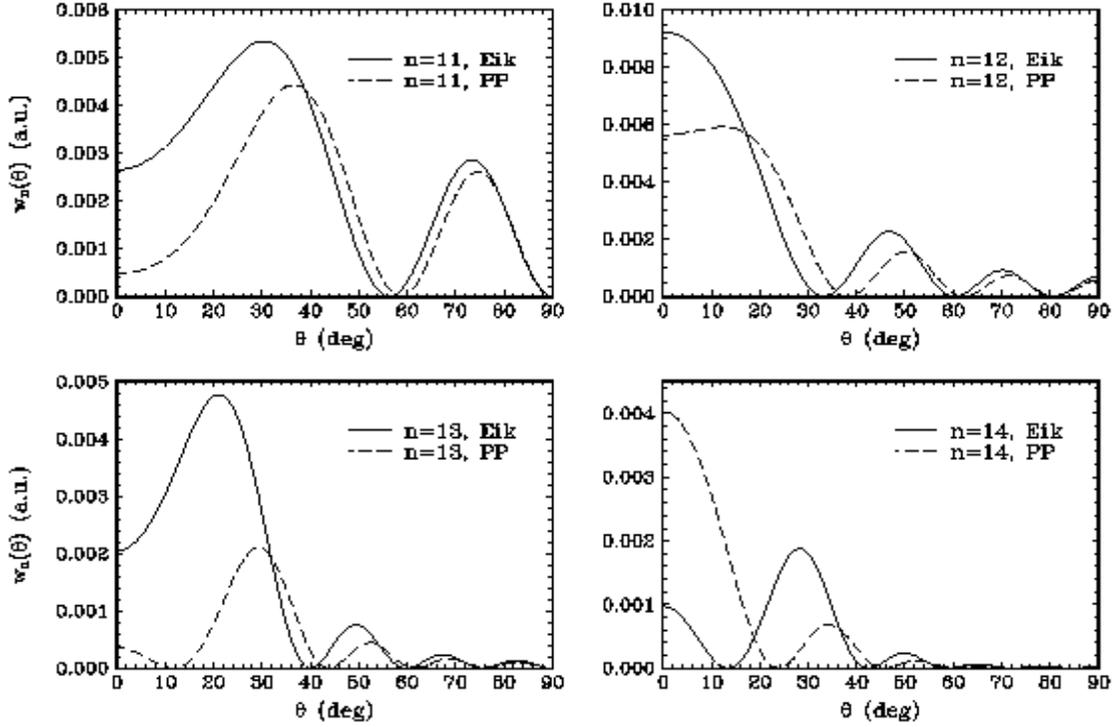,width=15cm,height=10cm} } 
\end{center}
\vspace{-2mm} 
\caption{ \small
Angle-resolved photoionization rates
$w_n(\theta)$ for four lowest open ATI channels
labeled by a number of absorbed photons $n$
($n=11$ corresponds to the lowest open ATI channel).
Hydrogen atom is illuminated by the laser wave with frequency
$\omega = 2$ eV and intensity $I=10^{14}$W/cm$^2$;
solid curves - eikonal theory, dashed curves -
the Perelomov-Popov theory \protect\cite{P3} (\protect\ref{coul}).}
\end{figure}
In order to address this issue
we develop the {\it eikonal}\/
approach for this problem. 
It  presents a simplified version of the semiclassical 
approximation, that assumes that the Coulomb field
does not produce significant distortions
of
classical trajectories
describing electron propagation in the continuum (i.e., the
electron wiggling motion in the laser field).
The Coulomb field comes into the picture through its 
contribution to the action $\int^t (Z/r) dt$ calculated 
along the classical trajectories discussed.
Within the semiclassical approximation the action plays a role of the wave
function phase.
It is important that the Coulomb field can produce large contribution
to the action while its distortion of the trajectories can remain
unsubstantial
Calculating the action, one is able to construct the
semiclassical wave function for the released electron,
and find with its help the ionization amplitude. 
If we totally neglect in this scheme
the Coulomb field putting $Z=0$, the eikonal 
wave function simplifies to be the Volkov function, 
and we return to the Keldysh-type approximation. An important 
verification of our eikonal approach 
provides the limit of low photoelectron energy where 
we reproduce the Perelomov-Popov result (\ref{coul}).
Fig.~6  presents a
quantitative
example which
illustrates importance of the Coulomb interaction.

\section{Conclusions}
\label{con}

Atomic antenna gives a clear physical idea how the complicated
multiphoton processes are operative. 
Absorption of large number of quanta from the laser field
needs that one of the atomic electrons is at first released
from an atom. After that, propagating in the core vicinity,
the electron accumulates high energy from the laser field and,
returning to the atomic core, transfers this energy into other
channels such as HHG, ATI or others.
Importantly, this physical picture
is implemented in a simple and reliable formalism.
We discussed two convenient ways to present the
theory.
One of them, called the factorization technique,
is presented by Eqs.~(\ref{sumc}), (\ref{A0A1}) for
the cases of HHG and ATI. In this approach the 
amplitude of a complicated process is expressed 
via the physical amplitudes of more simple,
"elementary" processes.
Another
scheme,
called the effective channels method,
is  based on Eq.~(\ref{s''}) for HHG. The effective channels
are closely related to the classical trajectories,
that makes them convenient for qualitative, as well as 
numerical studies.
We demonstrated an equivalence of the two approaches.

Applications of both formalisms need calculation of the
"elementary" amplitudes. This can be achieved by using
the modified Keldysh-type approach 
which very accurately reproduces
data for photodetachment, and, hopefully, for 
recombination and electron-atom scattering
in the laser field as well.
The Coulomb field of the atomic core can be taken into
account within eikonal approximation.
Reliability of the theoretical approaches is demonstrated
by quantitative applications.


\vspace{-2mm}
\begin{thebibliography}{99}
\bibitem{Ku87}
\vspace{-1mm}
M.~Yu.~Kuchiev, Pis'ma Zh. Eksp. Teor. Fiz. {\bf 45}, 319 (1987)
[JETP Letters {\bf 45}, 404 (1987)].

\bibitem{C93}
\vspace{-1mm}
P.~B.~Corkum, 
Phys. Rev. Lett. {\bf 71}, 1994 (1993).

\bibitem{KulanderA}
\vspace{-1mm}
J.~L.~Krause, K.~J.~Schafer, and K.~C.~Kulander, 
Phys. Rev. Lett. {\bf 68}, 3535 (1992).

\bibitem{KulanderB}
\vspace{-1mm}
K.~C.~Kulander, K.~J.~Schafer, and J.~L.~Krause,
in {\it Super-Intense 
Laser-Atom Physics}, Vol.~316 of {\it NATO Advanced
Study Institute, Series B: Physics}, edited by
B.~Piraux {\it et al} (Plenum, New York, 1993), p.~95.

\bibitem{K95}
\vspace{-1mm}
M.~Yu.~Kuchiev, J. Phys. B {\bf 28}, 5093 (1995).        

\bibitem{KO98let}
\vspace{-1mm}
M.~Yu.~Kuchiev and V.~N.~Ostrovsky, J. Phys. B 
{\bf 32}, L189 (1999).       

\bibitem{KO99pap}
\vspace{-1mm}
M.~Yu.~Kuchiev and V.~N.~Ostrovsky,  Phys. Rev. A {\bf 60}, 3111 (1999).     

\bibitem{Lew}
\vspace{-1mm} 
M.~Lewenstein, Ph.~Balcou, M.~Yu.~Ivanov, A.~L'Huillier, and P.~B.~Corkum,
Phys. Rev. A {\bf 49}, 2117 (1994).

\bibitem{Lewphase}
\vspace{-1mm}
M.~Lewenstein, P.~Salli\`{e}rs, and A.~L'Huillier, 
Phys. Rev. A {\bf 52}, 4747 (1995).

\bibitem{KO00eff}
\vspace{-1mm}
M.~Yu.~Kuchiev and V.~N.~Ostrovsky, http://xxx.lanl.gov/physics/0007016.
   
\bibitem{Keldysh}
\vspace{-1mm}
L.~V.~Keldysh, Zh.\'Eksp.Teor.Fiz.{\bf 47}, 1945 (1964)
[Sov.Phys.JETP {\bf 20}, 1307 (1965)].

\bibitem{P1}
\vspace{-1mm}
A.~M.~Perelomov, V.~S.~Popov, and M.~V.~Terent'ev, 
Zh. \'{E}ksp. Teor. Fiz. {\bf 50}, 1393 (1966)
[Sov. Phys.-JETP {\bf 23}, 924 (1966)].

\bibitem{P2}
\vspace{-1mm}
A.~M.~Perelomov, V.~S.~Popov, and M.~V.~Terent'ev, 
Zh. \'{E}ksp. Teor. Fiz. {\bf 51}, 309 (1966)
[Sov. Phys.-JETP {\bf 24}, 207 (1967)].

\bibitem{P3}
\vspace{-1mm}
A.~M.~Perelomov and V.~S.~Popov, 
Zh. \'{E}ksp. Teor. Fiz. {\bf 52}, 514 (1967)
[Sov. Phys.-JETP {\bf 25}, 336 (1967)].

\bibitem{GK97pra}
\vspace{-1mm}
G.~F.~Gribakin, M.~Yu.~Kuchiev, Phys. Rev. A {\bf 55}, 3760 (1997).

\bibitem{GK97jpb}
\vspace{-1mm}
G.~F.~Gribakin, M.~Yu.~Kuchiev, J. Phys. B {\bf 30}, 
L657 (1997).

\bibitem{KOdetach98}
\vspace{-1mm}
M.~Yu.~Kuchiev and V.~N.~Ostrovsky, J. Phys. B {\bf 31}, 2525 (1998).

\bibitem{KOdetach99}
\vspace{-1mm}
M.~Yu.~Kuchiev and V.~N.~Ostrovsky, Phys. Rev. A {\bf 59}, 2844 (1999).

\bibitem{T}
\vspace{-1mm}
D.~A.~Telnov, J.~Wang, and S.~I.~Chu, Phys. Rev. A {\bf 51}, 4797 (1995).

\bibitem{KO00}
\vspace{-1mm}
M.~Yu.~Kuchiev and V.~N.~Ostrovsky, Phys. Rev. A {\bf 61}, 033414 (2000).     

\bibitem{Jar}
\vspace{-1mm}
A.~Jaro\'{n}, J.~Z.~Kami\'{n}sky, and F.~Ehlotzky,
Phys. Rev. A {\bf 61}, 023404 (2000).

\bibitem{BLM94}  
\vspace{-1mm}
W.~Becker, S.~Long, and J.~K.~McIver, Phys. Rev. A {\bf 50}, 1540 (1994).


\bibitem{Chin}
\vspace{-1mm}
S.~F.~J.~Larochelle, A.~Talebpour, and S.~L.~Chin,
J. Phys. B {\bf 31}, 1215 (1998).

\bibitem{Pau}
\vspace{-1mm}
G.~G.~Paulus, W.~Becker, W.~Nicklich, and H.~Walter,
J. Phys. B {\bf 27}, L703 (1994).

\end{thebibliography}
\end{document}